\documentclass[fp,twocolumn]{jpsj3}
\usepackage{amsmath,amssymb,bm}
\usepackage{mathtools}
\usepackage{braket}
\usepackage{graphicx}
\usepackage[colorlinks=true,linkcolor=blue,citecolor=blue,urlcolor=blue]{hyperref}
\usepackage[T1]{fontenc}
\usepackage{newtxtext,newtxmath}
\usepackage{balance}
\allowdisplaybreaks

\makeatletter
\if@twocolumn\@lastpagebalancingfalse\fi
\providecommand{\ext@figure}{lof}
\providecommand{\ext@table}{lot}
\renewenvironment{acknowledgments}{\@acktrue\par\footnotesize\paragraph{Acknowledgment}}{\par}
\makeatother

\title{Compact Projected Pfaffian States for Rydberg-Blockaded Quantum Spin Systems}
\author{Ryui Kaneko\thanks{E-mail address: ryuikaneko@sophia.ac.jp}}
\inst{Physics Division, Sophia University, Chiyoda, Tokyo 102-8554, Japan}
\recdate{\today}

\abst{%
We numerically demonstrate that a compact projected Pfaffian ansatz effectively describes the ground state of the nonintegrable PXP model.
The variational wave function is obtained by applying separate Gutzwiller and Rydberg-blockade projectors to a general fermionic paired state,
and each blockade-allowed configuration has an amplitude given by the Pfaffian of a configuration-selected pairing matrix.
The exact blockade-projected product state of a known frustration-free deformation of the PXP model lies within the variational manifold
and provides a physically motivated initial state for the variational Monte Carlo optimization toward the ground state of the pure PXP model.
For a one-dimensional system of $N$ sites,
optimizing all symmetry-allowed pairings in an $O(N)$-parameter wave function yields
a near-exact ground state with a normalized energy variance (V-score) of order $10^{-9}$
despite the exponentially large blockade-constrained Hilbert space.
On the square lattice, an $O(N)$-parameter wave function with the same amplitude structure
yields relative energy errors below $0.2\%$
for clusters accessible to exact diagonalization
and produces magnetizations consistent with available benchmarks.
These results show that the same projected Pfaffian construction can be applied to blockade-constrained systems in both one and two dimensions.
}

\begin{document}

\maketitle

\section{Introduction}
\label{sec:introduction}

Strong local constraints can qualitatively alter the structure of quantum many-body wave functions.
Quantum simulators with Rydberg-atom arrays provide a programmable setting~\cite{saffman2010,browaeys2020}
in which strong interactions forbid simultaneous excitation of neighboring atoms,
leading to the PXP Hamiltonian in the blockade regime~\cite{labuhn2016,bernien2017,ebadi2021,scholl2021}.
Generic nonintegrable systems are expected to thermalize according to the eigenstate thermalization hypothesis~\cite{deutsch1991,srednicki1994}.
The nonintegrable PXP model is instead best known for quantum many-body scars~\cite{turner2018a,turner2018b,khemani2019,serbyn2021,moudgalya2022,park2025},
which hinder thermalization for special initial states
and produce long-lived revivals~\cite{bernien2017,choi2019,bluvstein2021}.
Their underlying structure has been elucidated through emergent approximate $\mathrm{SU}(2)$ algebras~\cite{choi2019,bull2020},
general projector-embedding constructions~\cite{shiraishi2017},
exact constrained eigenstates~\cite{lesanovsky2011,lesanovsky2012},
and frustration-free deformations~\cite{mark2020}.
These extensive studies of atypical excited states raise a complementary question about the ground state:
although no exact wave function is known for the pure PXP model,
can a compact and physically motivated variational family represent it accurately?

Variational constructions formulated directly in the constrained Hilbert space include low-bond-dimension matrix-product-state (MPS) manifolds.
Such MPS manifolds have been used to describe revival dynamics~\cite{ho2019,michailidis2020},
exact scar states and excitations~\cite{lin2019,surace2021},
connections to the Lesanovsky deformation~\cite{mark2020},
and volume-law scar states in one and higher dimensions~\cite{ivanov2025}.
With additional computational resources, tensor-network approaches can also describe higher-dimensional ground states~\cite{orus2014,li2024}.
These developments motivate representations that enforce the blockade exactly,
retain a common structure across dimensions, and admit systematic extensions.

Projected fermionic states provide such a complementary framework.
Previous fermionic treatments of the PXP model used auxiliary fermions and composite spins
to approximate the ground state and construct scar excitations~\cite{pan2022},
while scarred states in a longitudinal field
have been related to pairwise occupied Fermi-sea states near the Ising critical point~\cite{yao2022}.
More broadly, Gutzwiller-projected Bardeen--Cooper--Schrieffer (BCS) states encode hopping and pairing structures before projection:
Abrikosov fermions enlarge the local Hilbert space,
and Gutzwiller projection restores one fermion per site~\cite{anderson1987,gros1988,zhang1988,paramekanti2004,wen2002,becca2017}.
Projected fermionic states have been used as accurate variational descriptions~\cite{anderson1987,gros1988,paramekanti2004,lee2006,liu2012,kaneko2016,becca2017}
and also as exact ground-state representations for certain Hamiltonians~\cite{haldane1988,shastry1988,tu2013,misawa2026,kaneko2026}.
A BCS-type state has also been proposed for a Rydberg-blockaded quantum spin liquid on the kagome lattice~\cite{cheng2023},
motivated by a lattice-gauge-theory mapping~\cite{cheng2022};
there, however, auxiliary fermions are introduced to encode the blockade constraint,
rather than to provide a conventional Abrikosov-fermion representation of the physical spins~\cite{cheng2023,cheng2024}.

\begin{figure}[t]
  \centering
  \includegraphics[width=.75\columnwidth]{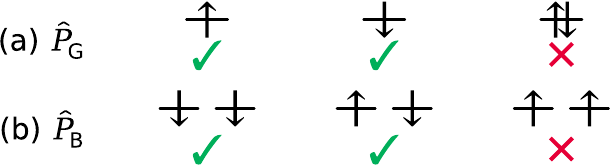}
  \caption{(Color online)
    Physical and Rydberg-blockade constraints.
    (a) The Gutzwiller projector $\hat{P}_{\mathrm G}$ selects the singly occupied local states.
    (b) The blockade projector $\hat{P}_{\mathrm B}$ removes nearest-neighbor $\ket{\uparrow,\uparrow}$ configurations from the singly occupied spin space.
    The two commuting projectors impose different constraints on the BCS state.
  }
  \label{fig:pgpb}
\end{figure}

Here we introduce a BCS ansatz with separate Gutzwiller and blockade projections (see Fig.~\ref{fig:pgpb})
for the ground state of the PXP model.
The amplitude of each allowed spin configuration is the Pfaffian
of the corresponding configuration-selected submatrix of a general two-component pairing matrix.
The construction enforces both single occupancy and the blockade exactly,
preserves the same amplitude structure in one and higher dimensions,
and remains compact because it contains only $O(N)$ symmetry-allowed pairing parameters,
where $N$ is the number of sites.
This linear parameter growth contrasts with the exponential growth of the blockade-constrained Hilbert space;
for a one-dimensional (1D) chain, its dimension scales asymptotically as $\varphi^N$,
with $\varphi=(1+\sqrt{5})/2$~\cite{li2024}.

We first embed the known frustration-free ground state into this projected BCS manifold.
Along the solvable line, the exact ground state is a blockade-projected product of single-site states~\cite{lesanovsky2011,lesanovsky2012,mark2020};
we show that it is equivalently a doubly projected BCS state whose pairing matrix factorizes into spatial and spin-dependent parts.
Although this exact correspondence applies only to the ground state along the frustration-free line,
it provides an analytically controlled starting point for studying the pure PXP Hamiltonian.
We then remove the factorization constraint
and optimize all pairing amplitudes by variational Monte Carlo (VMC).
In 1D, the resulting $O(N)$-parameter projected BCS state gives a near-exact ground state with a V-score of order $10^{-9}$.
On the two-dimensional (2D) square lattice,
the same configuration-amplitude structure remains tractable for system sizes beyond the reach of exact diagonalization (ED)
and yields energies and magnetizations consistent with available benchmarks.
These results establish projected fermionic pairing as a compact variational framework for blockade-constrained ground states in 1D and 2D.

The paper is organized as follows.
In Sect.~\ref{sec:hamiltonian}, we introduce the constrained Hamiltonian and its frustration-free deformation.
In Sect.~\ref{sec:ansatz}, we formulate the projected BCS ansatz,
establish its exact connection to the frustration-free ground state,
and specify the parameterizations in 1D and 2D.
In Sect.~\ref{sec:method}, we describe the VMC method and diagnostics.
In Sect.~\ref{sec:results}, we present results for the 1D chain and square lattice.
We summarize our results in Sect.~\ref{sec:conclusions}.

\section{PXP Model and Its Frustration-Free Deformation}
\label{sec:hamiltonian}

We first introduce the blockade-constrained spin Hamiltonian used throughout this work.
It contains the pure PXP model and a frustration-free model with an exactly known ground state as special cases.

We identify the local atomic ground and Rydberg-excited states with spin-$1/2$ states as
\begin{align}
  \ket{g_i}=\ket{0_i}=\ket{\downarrow_i},
  \quad
  \ket{r_i}=\ket{1_i}=\ket{\uparrow_i}.
  \label{eq:rydberg-spin-convention}
\end{align}
Here $\ket{g_i}$ and $\ket{r_i}$ denote the atomic ground and Rydberg-excited states at site $i$, respectively.
We take the lattice spacing to be unity and set the reduced Planck constant $\hbar=1$.
We use $x_i=0$ and $x_i=1$ to denote spin-down and spin-up states, respectively.
For spin operators, we use the convention
\begin{align}
  \hat{\nu}_{i}
  &= \hat{n}_{i,\uparrow}
  = \ket{\uparrow_i} \! \bra{\uparrow_i}
  = \frac{1+\hat{\sigma}_i^z}{2},
\\
  \hat{P}_{i}
  & =\hat{n}_{i,\downarrow}
  = 1-\hat{n}_{i,\uparrow}
  = \ket{\downarrow_i} \! \bra{\downarrow_i}
  = \frac{1-\hat{\sigma}_i^z}{2},
\\
  \hat{\sigma}_i^x
  &= \ket{\downarrow_i} \! \bra{\uparrow_i}
  + \ket{\uparrow_i} \! \bra{\downarrow_i},
\end{align}
satisfying $\hat{\sigma}_i^z\ket{\downarrow_i}=-\ket{\downarrow_i}$
and $\hat{\sigma}_i^z\ket{\uparrow_i}=+\ket{\uparrow_i}$.
We write $\mathrm{nn}(i)$ for the set of nearest neighbors of site $i$ and define
\begin{align}
  \hat{Q}_i & \coloneqq \prod_{j\in\mathrm{nn}(i)}\hat{P}_{j}=\prod_{j\in\mathrm{nn}(i)}\hat{n}_{j,\downarrow}, \\
  \hat{P}_{\mathrm B} & \coloneqq \prod_{\langle i,j\rangle} \left(1-\hat{\nu}_{i}\hat{\nu}_{j}\right)
  = \prod_{\langle i,j\rangle} \left(1-\hat{n}_{i,\uparrow}\hat{n}_{j,\uparrow}\right).
  \label{eq:blockade-projector}
\end{align}
For example, $\hat{Q}_i=\hat{P}_{i-1}\hat{P}_{i+1}=\hat{n}_{i-1,\downarrow}\hat{n}_{i+1,\downarrow}$ holds in 1D.
The Hamiltonian of the constrained model~\cite{lesanovsky2011,lesanovsky2012,mark2020} is defined by
\begin{align}
  \hat{H}_{\mathrm{constrained}}
   & \coloneqq \sum_i \hat{Q}_i\left(J_X\hat{\sigma}_i^x+J_n \hat{\nu}_{i} +J_P \hat{P}_{i}\right)\hat{Q}_i
  \\
   & = \sum_i \hat{Q}_i\left(J_X\hat{\sigma}_i^x+J_n \hat{n}_{i,\uparrow} +J_P \hat{n}_{i,\downarrow}\right)\hat{Q}_i.
  \label{eq:H_constrained}
\end{align}
One of the two factors of $\hat{Q}_i$ in each sandwiched local term in Eq.~\eqref{eq:H_constrained} is redundant
because $\hat{Q}_i$ commutes with the onsite operators ($\hat{\sigma}_i^x$, $\hat{\nu}_i$, and $\hat{P}_i$)
and is a projector ($\hat{Q}_i^2=\hat{Q}_i$).

For any real $z$ with $zJ_X>0$, the parameter line
\begin{align}
  J_n=\frac{J_X}{z},\quad J_P=zJ_X \label{eq:ffline}
\end{align}
is frustration-free~\cite{lesanovsky2011,lesanovsky2012,mark2020}.
Each local term is proportional to a positive projector onto a single local state:
$
  \hat{\sigma}_i^x + \hat{\nu}_i/z + z\hat{P}_i
  =
  \left(\ket{\uparrow_i}+z\ket{\downarrow_i}\right)
  \left(\bra{\uparrow_i}+z\bra{\downarrow_i}\right)
  / z
$.
Since $\ket{\downarrow_i}-z\ket{\uparrow_i}$ is orthogonal to $\ket{\uparrow_i}+z\ket{\downarrow_i}$,
the zero-energy ground state is
\begin{align}
  \ket{\Psi_{\rm FF}(z)} \propto \hat{P}_{\mathrm B} \bigotimes_i(\ket{\downarrow_i}-z\ket{\uparrow_i}).
  \label{eq:ffstate}
\end{align}
Its equal-time diagonal ground-state correlations map onto those of
a classical nearest-neighbor exclusion lattice gas
with statistical weight $z^{2N_\uparrow}$ for a configuration containing $N_\uparrow$ Rydberg excitations.
In 1D, no transition occurs at finite $z$,
and the quantum model is rigorously known to be gapped for $0<z<1$~\cite{katsura2013}.
On the square lattice, the corresponding classical hard-square model becomes critical
at $z_{\mathrm c}\simeq e^{1.334/2}\simeq 1.948$~\cite{guo2002}.

In the remainder of this work,
we consider the ground state of the pure PXP model~\cite{bernien2017,turner2018a,turner2018b},
obtained by setting $J_n=J_P=0$.
We set $J_X=1$, and the Hamiltonian becomes
\begin{align}
  \hat{H}
   =\sum_i \hat{Q}_i \hat{\sigma}_i^x \hat{Q}_i.
  \label{eq:H}
\end{align}
The pure PXP model belongs to the broader class of nearest-neighbor-exclusion or hard-boson models~\cite{fendley2004}.
Unlike the frustration-free deformation discussed above, the pure PXP Hamiltonian has no known closed-form ground-state wave function.
We treat the ground state of the pure PXP model variationally.
Hereafter, we focus on even system sizes $N$ under periodic boundary conditions and report energies in units of $J_X$.

\section{Gutzwiller- and Blockade-Projected BCS Ansatz}
\label{sec:ansatz}

We construct a fermionic BCS ansatz that satisfies both the single-occupancy and Rydberg-blockade constraints.
We first derive its configuration amplitudes and then establish its connection to the exact frustration-free ground state.

\subsection{Configuration amplitudes}

We represent the local spin states by Abrikosov fermions,
\begin{align}
  \ket{\downarrow_i}=\hat{c}_{i,\downarrow}^\dagger\ket{\mathrm{vac}}, \quad \ket{\uparrow_i}=\hat{c}_{i,\uparrow}^\dagger\ket{\mathrm{vac}},
\end{align}
where $\ket{\mathrm{vac}}$ is the fermionic Fock vacuum and is distinct from the local state $\ket{0_i}=\ket{\downarrow_i}$.
With $\hat{n}_{i,\sigma}=\hat{c}_{i,\sigma}^\dagger \hat{c}_{i,\sigma}$, 
the physical single-occupancy space is selected by
\begin{align}
  \hat{P}_{\mathrm G}
  \coloneqq
  \prod_i\left[ \hat{n}_{i,\uparrow}(1-\hat{n}_{i,\downarrow}) +(1-\hat{n}_{i,\uparrow})\hat{n}_{i,\downarrow}\right].
  \label{eq:gutzwiller-projector}
\end{align}
The Rydberg-blockade constraint is imposed separately by the projector $\hat{P}_{\mathrm B}$ defined in Eq.~\eqref{eq:blockade-projector}.
We define the variational state for a real antisymmetric $2N\times2N$ matrix $F$ as
\begin{align}
  \ket{\Psi(F)}
  \coloneqq
  \hat{P}_{\mathrm B}\hat{P}_{\mathrm G} \exp\!
  \left(\sum_{i,j,\sigma,\sigma'}\frac{F^{\sigma,\sigma'}_{i,j}}{2}\hat{c}_{i,\sigma}^\dagger \hat{c}_{j,\sigma'}^\dagger\right) \ket{\mathrm{vac}}.
  \label{eq:ansatz}
\end{align}
The exponential in Eq.~\eqref{eq:ansatz} has even fermion parity.
Consequently, its projection onto the physical sector with exactly one fermion per site is nonzero only when the number of sites $N$ is even.
We restrict the projected Pfaffian ansatz and VMC calculations to even-$N$ clusters.
For a physical configuration $\bm{x}=(x_0,\ldots,x_{N-1})$ on an even-$N$ cluster,
each $x_i=0,1$ specifies the occupied spin orbital $\downarrow,\uparrow$, respectively, at site $i$.
We use the spin-major ordering $(0,\downarrow),\ldots,(N-1,\downarrow),(0,\uparrow),\ldots,(N-1,\uparrow)$
and set the occupied-orbital index to $q_i=i$ for $x_i=0$ and $q_i=N+i$ for $x_i=1$.
The wave-function amplitude is
\begin{align}
  \Psi_F(\bm{x}) &= \langle\bm{x}|\Psi(F)\rangle = P_{\mathrm B}(\bm{x})\,\operatorname{Pf}F(\bm{x}),
  \label{eq:pfamp}
  \\
  F_{i,j}(\bm{x}) &= F_{q_i,q_j},
  \label{eq:pfamp_fqq}
\end{align}
where $F(\bm{x})$ is the $N\times N$ antisymmetric submatrix obtained from the full $2N\times 2N$ pairing matrix
by selecting one spin orbital per site according to $\bm{x}$, with the rows and columns ordered by site.
Here,
$P_{\mathrm B}(\bm{x})$ is the blockade projector, defined below in Eq.~\eqref{eq:blockade-projector-configuration},
and $\operatorname{Pf}$ is the Pfaffian.
For an even-dimensional antisymmetric matrix $X$ of size $N=2n$, we use the convention
\begin{align}
  \operatorname{Pf}X
  =\frac{1}{2^n n!}\sum_{\theta\in \mathfrak{S}_{2n}}\operatorname{sgn}(\theta)
  \prod_{k=1}^{n} X_{\theta(2k-1),\theta(2k)}.
  \label{eq:pfaffian-definition}
\end{align}
For an antisymmetric matrix $X$,
the Pfaffian satisfies $\bigl(\operatorname{Pf}X\bigr)^2=\det X$.
The Pfaffian incorporates fermionic antisymmetry,
whereas $\hat{P}_{\mathrm G}$ and $\hat{P}_{\mathrm B}$ impose the physical and blockade constraints exactly.

Pfaffian pairing wave functions provide a flexible extension of determinant-based fermionic trial states
and can be sampled efficiently in VMC~\cite{bajdich2006,bajdich2008,tahara2008,kurita2015,misawa2019}
with Gutzwiller and Rydberg-blockade projections.
The blockade projector
\begin{align}
  P_{\mathrm B}(\bm{x})
  \coloneqq \braket{\bm{x}|\hat{P}_{\mathrm B}|\bm{x}}
  =\prod_{\langle i,j\rangle}\left(1-x_i x_j\right)
  \label{eq:blockade-projector-configuration}
\end{align}
removes configurations
that contain at least one nearest-neighbor $\ket{\uparrow_i,\uparrow_j}$ ($x_i=x_j=1$) pair.
In practice, both projections are imposed by restricting the sampled configuration space:
we sample only blockaded configurations in the physical singly occupied basis,
which contains exactly one Abrikosov fermion at each site.
Consequently, $P_{\mathrm B}(\bm{x})=1$ and $P_{\mathrm G}(\bm{x})\coloneqq\braket{\bm{x}|\hat{P}_{\mathrm G}|\bm{x}}=1$
for every sampled configuration.
If configurations in the enlarged fermionic Fock space without the blockade constraint were considered instead,
the amplitude would explicitly contain $P_{\mathrm B}(\bm{x})P_{\mathrm G}(\bm{x})$.

\subsection{Choice of initial state: exact representation at the frustration-free point}
\label{sec:exactrepresentation}

We next show that the frustration-free ground state can be represented
exactly within the projected BCS variational family.
This representation provides an analytic benchmark
and serves as the initial state for the VMC optimization of the pure PXP ground state.

We first rewrite the frustration-free ground state in Eq.~\eqref{eq:ffstate}
and make its configuration amplitudes explicit.
We expand the product state as
\begin{align}
  \bigotimes_{i=0}^{N-1} \left( \ket{\downarrow_i}-z\ket{\uparrow_i} \right)
  = \sum_{\bm{x}\in\{0,1\}^{N}} (-z)^{N_\uparrow(\bm{x})} \ket{\bm{x}},
\end{align}
where $N_\uparrow(\bm{x})=\sum_i x_i$ counts the Rydberg-excited (equivalently, spin-up) sites.
We then have
\begin{align}
  \ket{\Psi_{\mathrm{FF}}(z)}
  &\propto
  \sum_{\bm{x}\in\{0,1\}^{N}}
  \left[
  \prod_{\langle i,j\rangle} \left( 1-\delta_{x_i,1}\delta_{x_j,1} \right)
  \right]
  (-z)^{N_\uparrow(\bm{x})} \ket{\bm{x}}.
  \label{eq:ffstate2}
\end{align}
Each blockade-allowed configuration $\bm{x}$ has an amplitude proportional
to $(-z)^{N_\uparrow(\bm{x})}$, whereas each forbidden configuration has zero amplitude.

We then embed this ground state into the variational manifold of the projected BCS states.
For an even periodic chain, we choose the factorized pairing matrix
\begin{align}
  F^{\sigma,\sigma'}_{i,j}=S_{i,j}v_\sigma v_{\sigma'}, \quad \sigma,\sigma'\in\{\downarrow,\uparrow\},
  \label{eq:exactF}
\end{align}
where $\bm{v}$ is the normalized vector given by
\begin{align}
  \bm{v}
  =
  \begin{pmatrix}
    v_{\downarrow} \\ v_{\uparrow}
  \end{pmatrix}
  =
  \frac{1}{\sqrt{1+z^2}}
  \begin{pmatrix}
    1 \\ -z
  \end{pmatrix}
  \label{eq:exactspinor}
\end{align}
in the internal spin basis $(\downarrow,\uparrow)$.
In 1D, we choose the spatial matrix $S$ as the real antisymmetric matrix defined by
\begin{align}
  S_{i,j} =
  \begin{cases}
    +1, & j>i,\\
     0, & j=i,\\
    -1, & j<i.
  \end{cases}
  \label{eq:signkernel}
\end{align}
For a binary configuration $\bm{x}=(x_0,\ldots,x_{N-1})$,
the spin state at each site selects the corresponding component of $\bm{v}$:
we set $v_{x_i}=v_\downarrow$ for $x_i=0$ and $v_{x_i}=v_\uparrow$ for $x_i=1$.
Likewise, the orbital index selected from the full pairing matrix is $q_i=i$ for $x_i=0$ and $q_i=N+i$ for $x_i=1$.
The resulting configuration-selected Pfaffian matrix has elements
\begin{align}
  F_{i,j}(\bm{x})=F_{q_i,q_j}=v_{x_i}S_{i,j}v_{x_j}.
  \label{eq:f_eq_vsv}
\end{align}
We collect the configuration-dependent internal-state factors into the diagonal matrix
$
  D(\bm{x})=\operatorname{diag}(v_{x_0},\ldots,v_{x_{N-1}})
$.
The configuration-selected pairing matrix factorizes as
\begin{align}
  F(\bm{x})=D(\bm{x})S D(\bm{x}).
  \label{eq:congruence}
\end{align}
For the ordering in Eq.~\eqref{eq:signkernel}, $\operatorname{Pf}S=1$ for even $N$~\cite{kaneko2026}.
Then, the identity $\operatorname{Pf}(B X B^{\mathsf{T}})=\det(B)\operatorname{Pf}(X)$ gives
\begin{align}
  \operatorname{Pf}F(\bm{x})
  & =\det D(\bm{x})\operatorname{Pf}S
  = \prod_{i=0}^{N-1} v_{x_i} \cdot 1
  = \frac{(-z)^{N_\uparrow(\bm{x})}}{(1+z^2)^{N/2}}.
  \label{eq:exactpfaffian}
\end{align}
For fixed $z$ and $N$, the factor $(1+z^2)^{-N/2}$ is independent of the configuration $\bm{x}$
and can be absorbed into the overall normalization of the wave function.
Since the blockade projector removes configurations with adjacent occupied sites,
the physical configuration amplitude in Eq.~\eqref{eq:pfamp} becomes
\begin{align}
  \Psi_F(\bm{x})
  \propto
  \prod_{\langle i,j\rangle} \left( 1-\delta_{x_i,1}\delta_{x_j,1} \right) (-z)^{N_\uparrow(\bm{x})}.
  \label{eq:exactconfiguration}
\end{align}
This is precisely the configuration amplitude of the blockade-projected product state in Eq.~\eqref{eq:ffstate2}.
Therefore,
\begin{align}
  \ket{\Psi(F)}
  &\propto \hat{P}_{\mathrm B} \bigotimes_i \left( \ket{\downarrow_i}-z\ket{\uparrow_i} \right)
  \propto \ket{\Psi_{\mathrm{FF}}(z)}.
  \label{eq:exactprojectedBCS}
\end{align}
Thus, the projected BCS state constructed above is an exact zero-energy ground state of the frustration-free Hamiltonian,
rather than a variational approximation.
Note that the same factor $(-1)^{N_\uparrow(\bm{x})}$ also appears
in the known sign structure of the pure PXP ground state~\cite{park2026};
after removing this factor, its amplitudes can be chosen strictly positive,
as guaranteed by the Perron--Frobenius theorem.

This exact representation provides the initial state for the VMC optimization.
We prepare the initial state using the matrix elements in Eq.~\eqref{eq:exactF}
and optimize all real pairing amplitudes allowed within the chosen symmetry and boundary-condition sector.
Consequently, neither the factorized form $S_{i,j}v_\sigma v_{\sigma'}$
nor the projected product amplitudes $(-z)^{N_\uparrow}$
are imposed on the optimized pure PXP wave function.

An analogous factorized construction applies to the square lattice.
We replace the 1D matrix $S$ by a fixed nonsingular real antisymmetric matrix $S_{\rm 2D}$
and retain the internal factor $\bm{v}\bm{v}^{\mathsf{T}}$.
The matrix $S_{\mathrm{2D}}$ is used only to construct the initial state
and does not constrain the subsequent optimization of $F^{\sigma,\sigma'}_{i,j}$.
For every configuration $\bm{x}$, the Pfaffian of $D(\bm{x})S_{\rm 2D}D(\bm{x})$ is proportional to $(-z)^{N_\uparrow(\bm{x})}$.
Any choice with $\operatorname{Pf}S_{\rm 2D}\ne0$ yields the same projected product initial state after normalization.
Although any such $S_{\mathrm{2D}}$ is admissible,
we use the specific sparse matrix in Eqs.~\eqref{eq:S_2d_definition_1} and \eqref{eq:S_2d_definition_2},
defined in Sect.~\ref{sec:twodimpar}.

We set the fugacity of the initial state to $z=0.65$ in all calculations.
In 1D, this choice is motivated by Ref.~[\citenum{mark2020}].
The squared overlap between the initial state and the PXP ground state was
found to be maximal near $z\simeq0.65$, reaching $0.977$ at $N=26$~\cite{mark2020}.
Although no analogous criterion has been established for the square lattice,
we use the same value solely for initialization.
The subsequent optimization allows the internal pairing channels to vary independently.
Thus, $z$ is not treated as an additional variational parameter.

Blockade-projected product states on the frustration-free line underlie the exact solution
of the Lesanovsky model~\cite{lesanovsky2011,lesanovsky2012,mark2020}.
As demonstrated above, this state can also be expressed as a doubly projected fermionic BCS state.
This representation embeds the exact state into a larger Pfaffian variational manifold applicable away from the solvable line.

\subsection{Short-range parent Hamiltonian for the initial state in 1D}
\label{sec:shortrangeparent}

Any fermionic Gaussian state admits a quadratic parent Hamiltonian,
although the locality of such a parent Hamiltonian is not automatically ensured.
In 1D, the initial state introduced above admits a particularly simple parent Hamiltonian
with only nearest-neighbor hopping and pairing terms,
despite the long-range wave-function pairing $S_{i,j}=\operatorname{sgn}(j-i)$.
The parent Bogoliubov--de Gennes (BdG) Hamiltonian is given by
\begin{align}
  \hat H_{\mathrm{BdG}}
  &=
  \sum_{i,\sigma,\sigma'}
  \biggl(
  t^{\sigma,\sigma'}
  \hat c_{i,\sigma}^{\dagger}\hat c_{i+1,\sigma'}
  +
  \Delta^{\sigma,\sigma'}
  \hat c_{i,\sigma}^{\dagger}\hat c_{i+1,\sigma'}^{\dagger}
  \nonumber
  \\
  &\quad 
  +\mathrm{H.c.}
  \biggr)
  +
  \sum_{i,\sigma,\sigma'}
  \epsilon^{\sigma,\sigma'}
  \hat c_{i,\sigma}^{\dagger}\hat c_{i,\sigma'},
  \label{eq:bdg}
\end{align}
where $t^{\sigma,\sigma'}$ are nearest-neighbor hopping amplitudes, $\Delta^{\sigma,\sigma'}$ are pairing amplitudes,
and $\epsilon^{\sigma,\sigma'}$ specifies the onsite single-particle terms,
including the chemical potential and spin-mixing amplitudes.
These parameters satisfy
\begin{align}
  \begin{pmatrix}
  t^{\downarrow,\downarrow} & t^{\downarrow,\uparrow}\\
  t^{\uparrow,\downarrow} & t^{\uparrow,\uparrow}
  \end{pmatrix}
  &=
  \begin{pmatrix}
  \Delta^{\downarrow,\downarrow} & \Delta^{\downarrow,\uparrow}\\
  \Delta^{\uparrow,\downarrow} & \Delta^{\uparrow,\uparrow}
  \end{pmatrix}
  =
  -\frac{2}{1+z^2}
  \begin{pmatrix}
  1&-z\\
  -z&z^2
  \end{pmatrix},
  \label{eq:bdg_t_delta}
\\
  \begin{pmatrix}
  \epsilon^{\downarrow,\downarrow} & \epsilon^{\downarrow,\uparrow}\\
  \epsilon^{\uparrow,\downarrow} & \epsilon^{\uparrow,\uparrow}
  \end{pmatrix}
  &=
  \frac{M}{1+z^2}
  \begin{pmatrix}
  z^2&z\\
  z&1
  \end{pmatrix},
  \label{eq:bdg_epsilon}
\end{align}
where $M>0$ is an arbitrary constant.

To see how this short-range parent Hamiltonian generates the long-range pairing matrix in Eq.~\eqref{eq:exactF},
we use the orthonormal vectors $\bm{v}=(1,-z)^{\mathsf T}/\sqrt{1+z^2}$
and $\bm{w}=(z,1)^{\mathsf T}/\sqrt{1+z^2}$ in the internal-state space.
They satisfy $\bm{v}^{\mathsf T}\bm{w}=0$.
The matrices in Eqs.~\eqref{eq:bdg_t_delta} and \eqref{eq:bdg_epsilon} can be written as
$-2\bm{v}\bm{v}^{\mathsf{T}}$ and $M\bm{w}\bm{w}^{\mathsf{T}}$, respectively.
Thus, only the mode $\hat{a}_i=\sum_{\sigma}v_{\sigma}\hat{c}_{i,\sigma}$ is paired,
while the orthogonal mode $\hat{b}_i=\sum_{\sigma}w_{\sigma}\hat{c}_{i,\sigma}$ has positive single-particle energy $M$
and is unoccupied in the corresponding BdG ground state.

For an even chain, we impose antiperiodic boundary conditions on the conventional Abrikosov fermions
and define the discrete Fourier transform as
$\hat{a}_j = \sum_k e^{ikj} \hat{a}_k / \sqrt{N}$
and
$\hat{b}_j = \sum_k e^{ikj} \hat{b}_k / \sqrt{N}$,
with $k = (2m+1)\pi/N$ and $m=0,1,\dots,N-1$.
In momentum space, the parent Hamiltonian is given by
\begin{align}
  \hat{H}_{\mathrm{BdG}}
  &=
  \frac{1}{2}\sum_k
  \begin{pmatrix}
    \hat{a}_k^\dagger & \hat{a}_{-k}
  \end{pmatrix}
  \begin{pmatrix}
    \xi_k & \Delta_k\\
    \Delta_k^* & -\xi_k
  \end{pmatrix}
  \begin{pmatrix}
    \hat{a}_k\\
    \hat{a}_{-k}^\dagger
  \end{pmatrix}
  + M\sum_k \hat{b}_k^\dagger\hat{b}_k,
  \label{eq:parent_BdG_momentum}
\end{align}
where
$\xi_k = -4\cos k$ and $\Delta_k = -4i\sin k$.
The quasiparticle energy is $E_k = \sqrt{\xi_k^2+|\Delta_k|^2} = 4$.
With the convention that the BCS state is written as
$\exp(\sum_k f_k \hat a_k^\dagger\hat a_{-k}^\dagger / 2)\ket{\mathrm{vac}}$,
the pairing function is
$f_k = \Delta_k / (\xi_k + E_k) = -i\cot(k/2)$.
Its discrete Fourier transform gives
$f_{i,j} = \sum_k e^{ik(j-i)}f_k/N = \operatorname{sgn}(j-i) = S_{i,j}$ for $i\not=j$ and $f_{i,i}=0$,
with the same sign structure appearing in the fermionic embedding
of the 1D transverse-field Ising model~\cite{kaneko2026}.
Upon restoring the internal-state components, we recover
\begin{align}
  \begin{pmatrix}
    F_{i,j}^{\downarrow,\downarrow} & F_{i,j}^{\downarrow,\uparrow} \\
    F_{i,j}^{\uparrow,\downarrow}   & F_{i,j}^{\uparrow,\uparrow}
  \end{pmatrix}
  =
  \frac{S_{i,j}}{1+z^2}
  \begin{pmatrix}
     1 & -z \\
    -z & z^2
  \end{pmatrix}
  =
  S_{i,j}\,\bm{v}\bm{v}^{\mathsf T},
  \label{eq:pairing_components}
\end{align}
in agreement with Eq.~\eqref{eq:exactF}.
Note that we directly optimize the pairing $F^{\sigma,\sigma'}_{i,j}$
rather than these parameters in the BdG Hamiltonian during the VMC optimization.

\subsection{Parameterization in 1D}
\label{sec:onedimpar}

On an even periodic chain, we parameterize the real antisymmetric pairing matrix $F$
using three sets of real, distance-dependent coefficients:
$f^{\uparrow,\uparrow}(d)$, $f^{\uparrow,\downarrow}(d)$, and $f^{\downarrow,\downarrow}(d)$.
We use the internal-state ordering $(\downarrow,\uparrow)$.
For sites $i$ and $j$, the corresponding matrix elements are
\begin{align}
  \begin{pmatrix}
    F^{\downarrow,\downarrow}_{i,j}
    &
    F^{\downarrow,\uparrow}_{i,j}
    \\
    F^{\uparrow,\downarrow}_{i,j}
    &
    F^{\uparrow,\uparrow}_{i,j}
  \end{pmatrix}
  =
  \begin{pmatrix}
    S_{i,j}
    f^{\downarrow,\downarrow}(d_{i,j})
    &
    \eta_{i,j}f^{\uparrow,\downarrow}(d_{i,j})
    \\
    -\eta_{j,i}f^{\uparrow,\downarrow}(d_{i,j})
    &
    S_{i,j}
    f^{\uparrow,\uparrow}(d_{i,j})
  \end{pmatrix},
  \label{eq:F_1d_parameterization}
  \\
  d_{i,j}
  =
  \min\bigl(\lvert i-j\rvert,N-\lvert i-j\rvert\bigr),
  \quad
  \eta_{i,j}
  =
  \begin{cases}
    +1, & j\geq i,\\
    -1, & j<i,
  \end{cases}
\end{align}
where $d_{i,j}$ is the shortest periodic distance
and $\eta_{i,j}$ defines the site-ordering convention.
The reversed elements are fixed by antisymmetry:
$
  F^{\sigma',\sigma}_{j,i}
  =
  -F^{\sigma,\sigma'}_{i,j}
$.
We optimize the same-spin coefficients
$f^{\uparrow,\uparrow}(d)$ and $f^{\downarrow,\downarrow}(d)$ independently for $d=1,\ldots,N/2$
and the mixed-spin coefficients
$f^{\uparrow,\downarrow}(d)$ for $d=0,\ldots,N/2$.
The resulting number of independent real coefficients is
\begin{align}
  N_{\mathrm{var}}^{\mathrm{1D}}
  =
  2 \cdot \frac{N}{2}
  +
  \left(\frac{N}{2}+1\right)
  =
  \frac{3N}{2}+1.
\end{align}
We remove the physically irrelevant overall scale by normalizing the parameter vector.
This parameterization has the form
\begin{align}
  F=\sum_p f_p M_p,
  \label{eq:F_basis_expansion}
\end{align}
where each fixed matrix $M_p$ represents one of the spin- and distance-resolved channels defined above.
The coefficients $f_p$ are independent real variational parameters.
We retain the explicit distance representation in 1D for clarity,
while we construct the corresponding symmetry-adapted basis directly in 2D.

\subsection{Parameterization in 2D}
\label{sec:twodimpar}

Although we use a real antisymmetric pairing ansatz similar to that in 1D,
we impose projective translation symmetry on the square lattice through an explicit matrix basis in 2D.
We consider a periodic $L_x\times L_y$ lattice with $N=L_xL_y$ and label its
sites by
\begin{align}
  \operatorname{ind}(x,y) \coloneqq x+L_x y,
  \quad
  0\leq x<L_x,\quad 0\leq y<L_y.
\end{align}
The spin-major orbital ordering is
\begin{align}
  \operatorname{orb}(\sigma,x,y)
  \coloneqq
  \begin{cases}
    \operatorname{ind}(x,y),   & \sigma=\downarrow,\\
    N+\operatorname{ind}(x,y), & \sigma=\uparrow.
  \end{cases}
  \label{eq:2d-orbital-ordering}
\end{align}
This convention agrees with the ordering used in
Eq.~\eqref{eq:pfamp}.

We expand the pairing matrix according to Eq.~\eqref{eq:F_basis_expansion}.
The fixed real matrices $M_p$ are antisymmetric
and invariant under the projective translations specified below.
We first define the translation matrices $T_x$ and $T_y$
through their nonzero elements,
\begin{align}
  (T_x)_{\operatorname{ind}(x+1,y),\operatorname{ind}(x,y)}
  & \coloneqq
  (-1)^y\tau_x^{\delta_{x,L_x-1}},
  \\
  (T_y)_{\operatorname{ind}(x,y+1),\operatorname{ind}(x,y)}
  & \coloneqq
  \tau_y^{\delta_{y,L_y-1}},
  \label{eq:projective-translations-2d}
\end{align}
where periodic boundary conditions are imposed on the destination coordinates.
The corresponding twist factors are applied only upon crossing a boundary.

The translation matrices defined above provide a projective representation
of the lattice translation group for the conventional Abrikosov fermions.
They satisfy
\begin{align}
  T_xT_y = -T_yT_x,
  \label{eq:projective_translation}
\end{align}
and thus $T_xT_yT_x^{-1}T_y^{-1}=-1$ holds.
The negative sign may be viewed as a $\mathbb{Z}_2$ gauge phase
associated with a $\pi$ flux through each elementary plaquette.
This flux characterizes the sign structure seen by the Abrikosov fermions
and does not represent a physical magnetic flux in the PXP model.

The choice of boundary twists is important
for constructing a nonsingular initial pairing matrix $S_{\mathrm{2D}}$ on a finite torus.
The factors $\tau_x,\tau_y=\pm1$ specify the Abrikosov-fermion boundary conditions
along the two directions of the torus.
We use $\tau_x=\tau_y=-1$ in all calculations.
This twist sector is invariant under the interchange $x\leftrightarrow y$
and allows us to construct a nonsingular antisymmetric matrix $S_{\mathrm{2D}}$, as described below.
By contrast, the fully periodic choice $\tau_x=\tau_y=+1$ gives $S_{\mathrm{2D}}$ a zero mode
and hence $\det S_{\mathrm{2D}}=0$.

The corresponding translations on the full spin-orbital space are given by
\begin{align}
  \mathcal{T}_\mu
  =
  \begin{pmatrix}
    T_\mu & 0\\
    0 & T_\mu
  \end{pmatrix},
  \quad \mu=x,y.
\end{align}
For each $\sigma,\sigma'\in\{\downarrow,\uparrow\}$ and each oriented
displacement
\begin{align}
  \bm{\delta}=(\Delta x,\Delta y),
  \quad
  0\leq\Delta x<L_x,\quad 0\leq\Delta y<L_y,
\end{align}
we define the antisymmetric matrix
\begin{align}
  C_{\sigma,\sigma'}(\bm{\delta})
  &=
  \lvert \operatorname{orb}(\sigma,0,0)\rangle
  \langle \operatorname{orb}(\sigma',\Delta x,\Delta y)\rvert
  \nonumber\\
  &\quad-
  \lvert \operatorname{orb}(\sigma',\Delta x,\Delta y)\rangle
  \langle \operatorname{orb}(\sigma,0,0)\rvert.
\end{align}
We omit the case in which the two orbitals coincide,
i.e., $\sigma=\sigma'$ and $\bm{\delta}=(0,0)$.
We then average each matrix over the projective translations:
\begin{align}
  \widetilde{M}_{\sigma,\sigma'}(\bm{\delta})
  =\frac{1}{N}
  \sum_{n_y=0}^{L_y-1}
  \sum_{n_x=0}^{L_x-1}
  \mathcal{T}_x^{n_x}\mathcal{T}_y^{n_y}
  C_{\sigma,\sigma'}(\bm{\delta})
  \left(\mathcal{T}_x^{n_x}\mathcal{T}_y^{n_y}\right)^{\mathsf{T}}.
  \label{eq:M_2d_definition}
\end{align}
Each candidate matrix constructed in this way satisfies
$
  \mathcal{T}_{\mu} \widetilde{M}_{\sigma,\sigma'}(\bm{\delta}) \mathcal{T}_{\mu}^{\mathsf{T}}
  = \widetilde{M}_{\sigma,\sigma'}(\bm{\delta})
$
and
$
  \widetilde{M}_{\sigma,\sigma'}(\bm{\delta})^{\mathsf{T}} 
  = -\widetilde{M}_{\sigma,\sigma'}(\bm{\delta})
$.
Thus, the construction ensures
$
  \mathcal{T}_{\mu}F\mathcal{T}_{\mu}^{\mathsf{T}}=F
$
and
$
  F^{\mathsf{T}}=-F
$.

Note that the projective sign structure is a property of the Abrikosov-fermion description
and does not break the physical lattice translation symmetry of the projected wave function.
The resulting projected state transforms under the ordinary lattice translation group
and carries a well-defined crystal momentum.
Indeed, an auxiliary translation may be viewed as the combination $T_\mu=G_\mu P_\mu$,
where $P_\mu$ translates the lattice sites and $G_\mu$ is a site-dependent sign transformation.
After Gutzwiller projection, every site is occupied by exactly one fermion,
and hence $G_\mu$ contributes only an overall phase that is independent of the spin configuration.
Since the blockade projection commutes with the physical lattice translations,
it preserves the translation quantum number of the Gutzwiller-projected state.
Thus, the anticommutation relation in Eq.~\eqref{eq:projective_translation}
characterizes only the Abrikosov-fermion representation
and does not imply that the physical lattice translations anticommute.
Accordingly, we regard the present choice of $(\tau_x,\tau_y)=(-1,-1)$
primarily as a convenient finite-size boundary condition for the Abrikosov-fermion parameterization,
rather than as the selection of a distinct physical topological sector.

The candidate set may contain zero matrices and linear dependencies.
For instance, for $L\equiv0\pmod4$,
the same-spin candidate at $\bm{\delta}=(L/2,L/2)$ vanishes
because $-\bm{\delta}\equiv\bm{\delta}$
and the two contributions enter with opposite signs.
In addition, $\widetilde{M}_{\uparrow,\downarrow}(\bm{\delta})$
and $\widetilde{M}_{\downarrow,\uparrow}(-\bm{\delta})$ are linearly dependent.
To extract a linearly independent basis,
we consider the candidate matrices sequentially in the order
$\sigma=\downarrow,\uparrow$,
$\sigma'=\downarrow,\uparrow$,
$\Delta y=0,\ldots,L_y-1$,
and $\Delta x=0,\ldots,L_x-1$.
Candidates with a Frobenius norm smaller than $10^{-12}$ are discarded.
Each remaining candidate is normalized to unit Frobenius norm
and tested for linear independence against the previously retained directions by Gram--Schmidt orthogonalization.
If the residual norm exceeds $10^{-12}$,
the normalized candidate itself, rather than its orthogonalized residual, is retained and labeled $M_p$.
The resulting ordered basis fixes both the variational subspace and the parameter coordinates used in the calculation.

For the $L^2$ clusters with even $L$,
the rank of the candidate set (the number of independent variational parameters) is
\begin{align}
  N_{\mathrm{var}}^{\mathrm{2D}}
  =
  \begin{cases}
    2N,   & L\equiv0 \pmod 4,\\
    2N+2, & L\equiv2 \pmod 4,
  \end{cases}
  \quad N=L^2.
\end{align}
As in 1D, we remove the physically irrelevant overall scale by normalizing the parameter vector.

We initialize the optimization with a factorized pairing matrix in Eq.~\eqref{eq:f_eq_vsv}.
As one possible choice,
we take the factor $S_{\mathrm{2D}}$ to be a real nonsingular antisymmetric matrix
with nonzero elements assigned to positively oriented nearest-neighbor bonds:
\begin{align}
  \left(S_{\mathrm{2D}}\right)_{\operatorname{ind}(x,y),\operatorname{ind}(x+1,y)}
  & \coloneqq
  \tau_x^{\delta_{x,L_x-1}},
  \label{eq:S_2d_definition_1}
  \\
  \left(S_{\mathrm{2D}}\right)_{\operatorname{ind}(x,y),\operatorname{ind}(x,y+1)}
  & \coloneqq
  (-1)^x\tau_y^{\delta_{y,L_y-1}}.
  \label{eq:S_2d_definition_2}
\end{align}
We connect the nearest-neighbor bonds across the lattice boundaries using the specified twist factors.
The reverse-bond matrix elements follow from antisymmetry, and all non-nearest-neighbor elements vanish.
This sign assignment is locally consistent with a square-lattice Kasteleyn orientation~\cite{kasteleyn1963},
supplemented by twisted boundary bonds on the torus.

The initial matrix in the spin-major ordering is
\begin{align}
  F_0
  =
  \left(\bm{v}\bm{v}^{\mathsf{T}}\right)
  \otimes S_{\mathrm{2D}},
  \label{eq:F0_2d}
\end{align}
where $\bm{v}$ is defined in Eq.~\eqref{eq:exactspinor}.
Its matrix elements are explicitly given by
$
  (F_0)_{\sigma N+i,\sigma' N+j}
  =v_{\sigma} v_{\sigma'}(S_{\mathrm{2D}})_{i,j}
$,
where we identify $\downarrow$ and $\uparrow$ with block indices $0$ and $1$, respectively.
Since $F_0$ belongs to the variational subspace spanned by the retained basis matrices,
the expansion $F_0=\sum_p f_p^{(0)}M_p$ is exact up to numerical precision.
We obtain the initial coefficients $f_p^{(0)}$ by solving the corresponding linear least-squares problem
and then normalize the coefficient vector.
During the VMC optimization, all retained coefficients $f_p$ vary independently.
Therefore, the optimized pairing matrix is not restricted to the factorized form in Eq.~\eqref{eq:F0_2d}
or to nearest-neighbor pairing.

\section{Method}
\label{sec:method}

\subsection{VMC method}
\label{sec:vmc}

In the VMC calculations,
we sample configurations in the blockaded Hilbert space with probability proportional to $|\Psi_F(\bm{x})|^2$.
In our implementation, we do not recompute the factor $P_{\mathrm B}(\bm{x})$ for each Pfaffian evaluation.
Instead, we select a site uniformly from all $N$ sites and immediately reject the proposed flip
if it violates the blockade constraint.
An allowed flip is accepted with probability $\min(1,\lvert R_i\rvert^2)$,
defined below in Eq.~\eqref{eq:pfaffian-fast-ratio}.
Every configuration visited by the Markov chain satisfies $P_{\mathrm B}(\bm{x})=1$,
whereas configurations with $P_{\mathrm B}(\bm{x})=0$ are never generated.
The sampled amplitude is $\operatorname{Pf}F(\bm{x})$ within this restricted configuration space.
The blockade projector does not modify the Pfaffian ratio between two configurations connected by an allowed move.

We apply the Pfaffian and inverse-matrix fast-update strategies that are often used in
fermionic VMC implementations~\cite{bajdich2006,bajdich2008,tahara2008,morita2015,xu2022,chen2026}.
A single-site flip at site $i$ changes only the selected orbital $q_i$
and replaces the $i$th row and the corresponding column of the antisymmetric matrix $F(\bm{x})$.
In what follows, we omit the explicit dependence on $\bm{x}$ and write $F \coloneqq F(\bm{x})$ for notational simplicity.
Under the proposed move,
the antisymmetric matrix $F'$ is constructed by replacing the $i$th row of $F$ together with its corresponding column.
We write the updated $i$th row of $F'$ as the transpose of a column vector $\bm{r}^{(i)}$,
whose components are given by $r_j^{(i)} = F'_{i,j}$ for $j=0,\ldots,N-1$.
In particular, $r_i^{(i)}=F'_{i,i}=0$ because $F'$ is antisymmetric.
We define the corresponding change in the $i$th row by $\delta\bm{r}^{(i)}$,
with components $\delta r_j^{(i)} = r_j^{(i)}-F_{i,j}$ ($j=0,\ldots,N-1$).
For the $i$th Cartesian unit vector $\bm{e}_i$, satisfying $(\bm{e}_i)_j=\delta_{i,j}$,
the fully updated matrix $F'$ can be written as
\begin{align}
  F' = F +\bm{e}_i\delta\bm{r}^{(i)\mathsf{T}} -\delta\bm{r}^{(i)}\bm{e}_i^{\mathsf{T}}.
  \label{eq:pfaffian-rank-two-update}
\end{align}
Then, the Pfaffian ratio~\cite{tahara2008,morita2015,chen2026} is
\begin{align}
  R_i \coloneqq \frac{\operatorname{Pf}F'}{\operatorname{Pf}F} = 1+\delta\bm{r}^{(i)\mathsf{T}}F^{-1}\bm{e}_i
  = \bm{r}^{(i)\mathsf{T}}F^{-1}\bm{e}_i.
  \label{eq:pfaffian-fast-ratio}
\end{align}
The Metropolis acceptance probability for an allowed flip is $\min(1,\lvert R_i\rvert^2)$.
For an accepted move,
with $\bm{u}^{(i)}\coloneqq F^{-1}\bm{e}_i$ and $\bm{v}^{(i)}\coloneqq F^{-1}\delta\bm{r}^{(i)}$,
the inverse matrix is updated by
\begin{align}
  F'^{-1} = F^{-1} + R_i^{-1} ( \bm{u}^{(i)}\bm{v}^{(i)\mathsf{T}} - \bm{v}^{(i)}\bm{u}^{(i)\mathsf{T}} ).
  \label{eq:pfaffian-fast-inverse-update}
\end{align}
Note that
$\bm{v}^{(i)\mathsf{T}}=-\delta\bm{r}^{(i)\mathsf{T}}F^{-1}$ and $\bm{u}^{(i)\mathsf{T}}=-\bm{e}_i^{\mathsf{T}} F^{-1}$ because $F^{-1}$ is antisymmetric.
The matrix-vector product $F^{-1}\delta\bm{r}^{(i)}$ and the two outer products in Eq.~\eqref{eq:pfaffian-fast-inverse-update} require $O(N^2)$ operations.
By contrast, the ratio in Eq.~\eqref{eq:pfaffian-fast-ratio} requires only $O(N)$ operations
because $F^{-1}\bm{e}_i$ is the $i$th column of the stored inverse matrix.
The selected matrix and its inverse are recomputed at a cost of $O(N^3)$ after a sufficiently large number of iterations,
or whenever $R_i$ falls outside a physically reasonable range.
Thus, the explicit blockade projection ensures that the Monte Carlo (MC) chain samples only physically allowed configurations
while retaining the standard fast-update scheme for Pfaffian wave functions.
For the general constrained Hamiltonian in Eq.~\eqref{eq:H_constrained},
the local energy contains the diagonal contributions from the $J_n$ and $J_P$ terms
and a sum over the amplitude ratios for spin flips that preserve the blockade constraint.
Forbidden flips lead to configurations with zero wave-function amplitude and do not contribute to the sum.

For the pure PXP model, all symmetry-allowed real coefficients in the chosen pairing basis are treated as variational parameters
and optimized by the stochastic reconfiguration (SR) method~\cite{sorella1998,sorella2001,tahara2008},
which is equivalent to the natural gradient method~\cite{amari1998,nomura2017}.
The parameter update $\delta f_q$ at each iteration is obtained by solving
\begin{align}
  \sum_q(\mathcal{S}_{pq}+\lambda_p\delta_{pq})\,\delta f_q=-g_p,
  \label{eq:sr_method}
\end{align}
where $\mathcal{S}_{pq}=\langle\Delta O_p\Delta O_q\rangle$, $g_p=2\langle\Delta O_p\Delta E_{\rm loc}\rangle$, and $O_p=\partial_{f_p}\ln\Psi_F$.
Here, $\lambda_p>0$ is a diagonal regularization parameter introduced to stabilize the inversion of the SR matrix.
We further impose a bound on the norm of the parameter update.
We obtain the reported state by averaging the normalized parameters over the final optimization iterations.

\subsection{Numerical implementation}
\label{sec:num_imp}

We performed all calculations with the pure PXP couplings $J_X=1$ and $J_n=J_P=0$.
The variational parameters were initialized from the frustration-free state at $z=0.65$.
To break possible accidental symmetries,
independent Gaussian noise with a standard deviation of $10^{-6}$ was added to the symmetry-reduced coefficients before normalization.

For the 1D systems, the optimization was carried out for $4000$ iterations using $5120$ MC samples per iteration and a learning rate of $0.05$.
For the 2D systems, we used $6000$ iterations and $10240$ samples per iteration.
In both cases, the diagonal shift in Eq.~\eqref{eq:sr_method} at iteration $t$ was set to $\lambda_t = 10^{-3}\max(0.1,0.95^t)$.
The normalized parameter vectors obtained over the final $100$ iterations were averaged and normalized to determine the final variational state.

The calculations were parallelized across $1024$ MPI ranks.
Every MPI rank performed $400$ and $500$ re-equilibration sweeps for the 1D and 2D systems, respectively,
before collecting samples at each optimization iteration.
For the final measurements, these numbers were increased to $1000$ and $2000$ thermalization sweeps, respectively.
A total of $204800$ additional samples were generated, with one sample recorded after each MC sweep.
To control the accumulation of numerical errors in the fast-update procedure,
the current Pfaffian matrix was reconstructed after every $256$ accepted moves,
or earlier if the rank-two inverse update in Eq.~\eqref{eq:pfaffian-fast-inverse-update} was found to be numerically unstable.
The statistical errors in the final measurements were estimated by a blocking analysis with $500$ gathered samples per block.

\subsection{Physical quantities}
\label{sec:phys_quant}

We use the following dimensionless measure of wave-function quality:
\begin{align}
  V
  \coloneqq
  N\frac{\langle \hat{H}^2\rangle-\langle \hat{H}\rangle^2} {(\langle \hat{H}\rangle-E_\infty)^2},
  \label{eq:vscore}
\end{align}
where $E_\infty$ is the infinite-temperature energy in the constrained Hilbert space.
For the pure PXP Hamiltonian, $E_\infty=0$
because $\langle\bm{x}|\hat{H}|\bm{x}\rangle=0$ for every constrained-basis state $|\bm{x}\rangle$.
The numerator $\langle \hat{H}^2\rangle-\langle \hat{H}\rangle^2$ in Eq.~\eqref{eq:vscore} is the energy variance,
whereas the denominator provides the corresponding squared energy scale;
the additional factor of $N$ removes the leading system-size dependence for extensive Hamiltonians.
Therefore, the V-score is an energy-scale-normalized variance and is dimensionless.
It can be evaluated without knowing the exact ground-state energy or wave-function overlap~\cite{wu2024}.
Benchmarks across many Hamiltonians and variational methods show that
it provides an empirical estimate of the order of magnitude of the relative energy error,
while also permitting comparisons in regimes where the ED result is unavailable~\cite{wu2024}.
The empirical relation is
\begin{align}
  \ln \epsilon_E = \ln V + C,\quad C=-1.80\pm0.08,
  \label{eq:vscore-energy-error}
\end{align}
where $\epsilon_E=\lvert\langle \hat{H}\rangle-E_0\rvert/\lvert E_0-E_\infty\rvert$ is the relative energy error~\cite{wu2024}.
This relation gives $\epsilon_E\simeq e^C V\simeq0.165V$,
with a multiplicative uncertainty of approximately $8\%$
arising from the fitted intercept.
A nonzero V-score indicates that the optimized state is not an exact eigenstate.

We use the relation in Eq.~\eqref{eq:vscore-energy-error} only as a conditional order-of-magnitude reference.
Its quantitative applicability to the present blockade-constrained Hilbert space has not been independently established,
and the inferred values should not be interpreted as rigorous error bounds.
Wherever ED is available, we assess the accuracy directly from the energy difference and wave-function fidelity
rather than from this empirical conversion.
The fidelity is evaluated deterministically by enumerating the full constrained basis.

Hereafter, we show the ground-state energy density
\begin{align}
  e_{\mathrm{GS}} \coloneqq \frac{\langle \hat{H}\rangle}{N},
  \label{eq:energy-density}
\end{align}
and the magnitude of the longitudinal magnetization
\begin{align}
  m_z \coloneqq \left\lvert\frac{1}{N}\sum_i\langle\hat{\sigma}_i^z\rangle\right\rvert.
  \label{eq:longitudinal-magnetization}
\end{align}
The absolute value removes the dependence of the reported longitudinal magnetization on the overall sign convention for the spin-$z$ axis.
For the convention in Eq.~\eqref{eq:rydberg-spin-convention}, the Rydberg-excitation (hard-core-boson) density is
\begin{align}
  n
  \coloneqq \frac{1}{N}\sum_i\langle \hat{\nu}_i\rangle
  = \frac{1}{2}\left(1+\frac{1}{N}\sum_i\langle\hat{\sigma}_i^z\rangle\right).
  \label{eq:density-magnetization-conversion}
\end{align}
This becomes $n=(1-m_z)/2$ since the square-lattice ground state obtained here has negative uniform $z$ polarization.
One can also define the constrained transverse magnetization
$m_x=\left|\sum_i\langle \hat{Q}_i\hat{\sigma}_i^x\hat{Q}_i\rangle\right|/N$;
however, we show the energy density instead of the transverse magnetization
as $m_x=|e_{\mathrm{GS}}/J_X|$ holds for the pure PXP model ($J_n=J_P=0$).

\section{Results}
\label{sec:results}

\subsection{1D chain}

\begin{figure}[t]
  \includegraphics[width=\columnwidth]{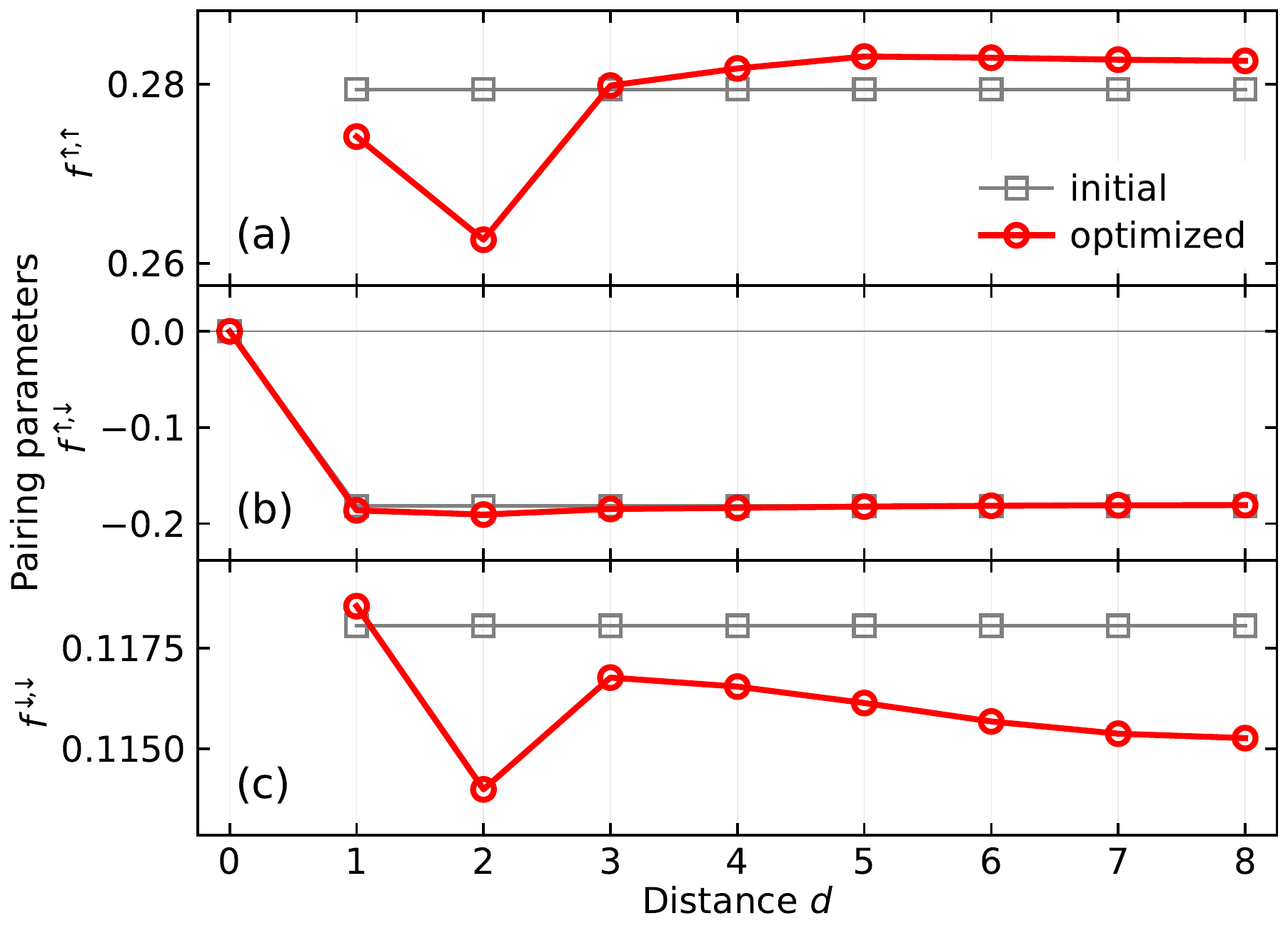}
  \caption{(Color online)
    Initial and optimized normalized pairing parameters for the 1D pure PXP chain at $N=16$.
    The three panels show the symmetry-reduced pairing amplitudes
    $f^{\uparrow,\uparrow}(d)$ (a), $f^{\uparrow,\downarrow}(d)$ (b), and $f^{\downarrow,\downarrow}(d)$ (c)
    as functions of the shortest periodic distance $d$.
    Gray squares denote the frustration-free projected product initial state with $z=0.65$,
    and red circles denote the parameters after optimization for the pure PXP Hamiltonian.
    The optimized parameters remain close to their initial values over all pairing channels and distances,
    indicating that the analytically motivated initial state already lies near the accurate variational solution,
    although the optimization changes the state nontrivially.
  }
  \label{fig:pairing-parameters-L16}
\end{figure}

\begin{figure}[t]
  \includegraphics[width=\columnwidth]{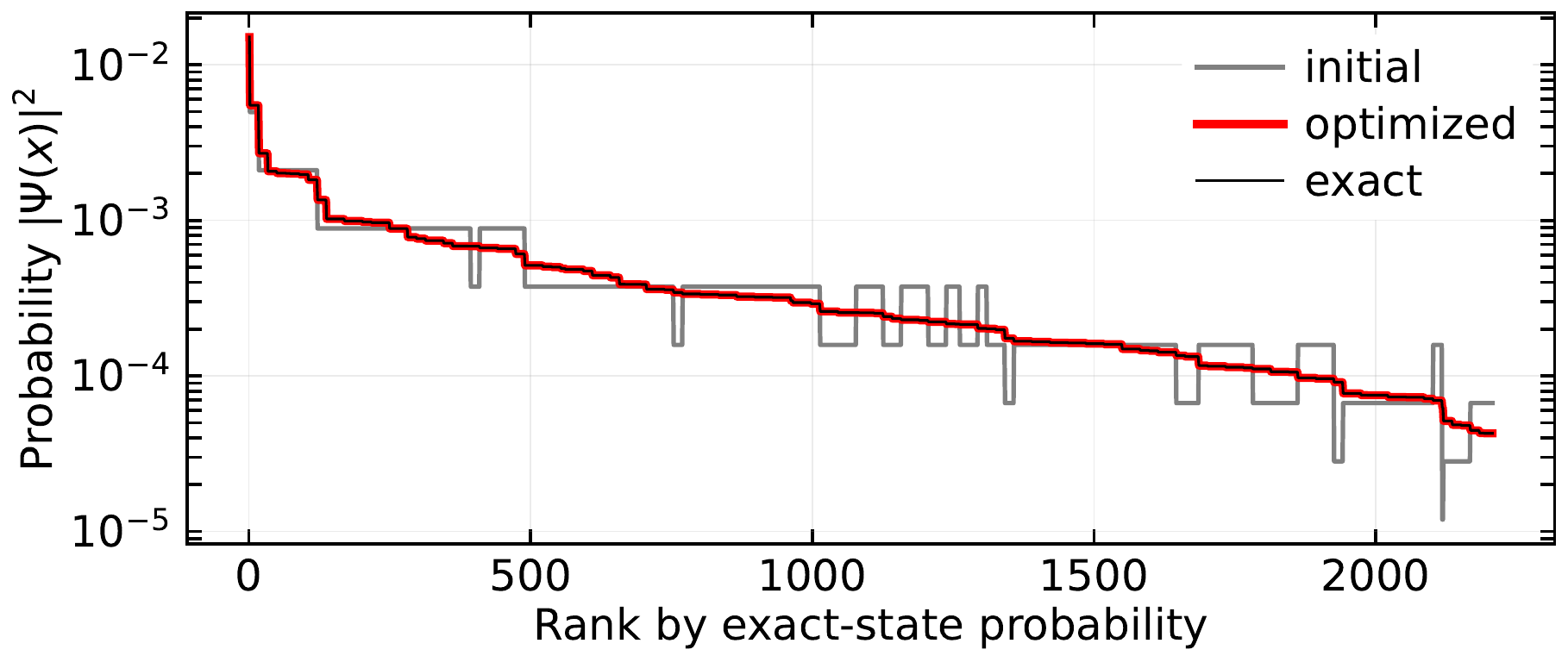}
  \caption{(Color online)
    Configuration probabilities for the 1D pure PXP chain at $N=16$.
    For every blockade-allowed binary configuration $\bm{x}$,
    the squared normalized amplitudes $\lvert\Psi(\bm{x})\rvert^2$ of the initial projected BCS state (gray line),
    the optimized projected BCS state (thick red line), and the ED ground state (thin black line) are shown.
    The horizontal axis shows the complete constrained spin basis sorted by the exact ground-state probability in descending order.
    The optimized distribution is nearly indistinguishable from the exact distribution over the full probability range,
    while the initial state captures the overall hierarchy with visible configuration-resolved deviations.
  }
  \label{fig:wavefunction-L16}
\end{figure}

\begin{table}[t]
  \centering
  \caption{Comparison between ED and VMC results for the energy density, V-score, and squared fidelity.
    The uncertainties in the VMC energy density and V-score indicate MC statistical errors.}
  \label{tab:energy-vscore-fidelity}
  \begin{tabular}{@{}c@{\,}c@{\,}c@{\!\!\!}c@{\!\!\!}c@{}}
    \hline
    $N$ & $e_{\mathrm{ED}}$ & $e_{\mathrm{VMC}}$ & $10^{10} \times \text{V-score}$ & Fidelity \\
    \hline
    \hline
    16 & $-0.6035618315$ & $-0.6035618058(84)$ & $6.710(31)$ & $0.999999999913$ \\
    18 & $-0.6035609334$ & $-0.6035609370(85)$ & $6.770(32)$ & $0.999999999897$ \\
    20 & $-0.6035606881$ & $-0.6035606891(76)$ & $6.714(35)$ & $0.999999999884$ \\
    22 & $-0.6035606197$ & $-0.6035606199(78)$ & $6.561(32)$ & $0.999999999877$ \\
    24 & $-0.6035606003$ & $-0.6035605919(71)$ & $6.293(30)$ & $0.999999999871$ \\
    26 & $-0.6035605947$ & $-0.6035606091(64)$ & $6.047(25)$ & $0.999999999865$ \\
    \hline
  \end{tabular}
\end{table}

To gain insight into the effect of optimization on the initial state,
we first compare the symmetry-reduced pairing parameters before and after optimization at $N=16$ in Fig.~\ref{fig:pairing-parameters-L16}.
The optimized parameters closely track the initial values in all three spin channels.
The largest visible changes occur mostly at a few short distances,
while the overall distance dependence is nearly preserved.
This agreement indicates that the frustration-free projected product state is more than a convenient nonsingular initial state:
it already provides a good starting approximation to the 1D pure PXP ground state.
At the same time, the closeness in parameter space alone does not guarantee the similarity of normalized many-body wave functions.

We next compare the optimized wave-function amplitudes with the exact results for $N=16$, as shown in Fig.~\ref{fig:wavefunction-L16}.
Because the physical amplitudes depend nonlinearly on the Pfaffians and the subsequent projections,
this comparison provides a direct assessment of the resulting many-body wave function.
For visualization, the basis configurations are sorted in descending order of their probabilities in the ED ground state,
and the same ordering is used for both the initial and optimized states.
The initial state captures the overall envelope of the exact probability distribution,
consistent with the relatively small changes in the pairing parameters shown in Fig.~\ref{fig:pairing-parameters-L16}.
Nevertheless, it exhibits step-like structures and configuration-dependent deviations from the exact result.
After optimization, the variational probabilities agree with the exact values.
This high level of agreement is also reflected in the squared fidelity with the ED ground state.
For the largest system treated by ED in this work, $N=26$, the squared fidelity reaches
$\mathcal{F}=\lvert\langle\Psi_{\rm ED}|\Psi_{\rm VMC}\rangle\rvert^2 = 0.999999999865$,
as reported in Table~\ref{tab:energy-vscore-fidelity}.
These results show that the independent optimization of the pairing channels removes the residual configuration-dependent deviations
that are present in the physically motivated initial state.

\begin{figure}[t]
  \centering
  \includegraphics[width=\columnwidth]{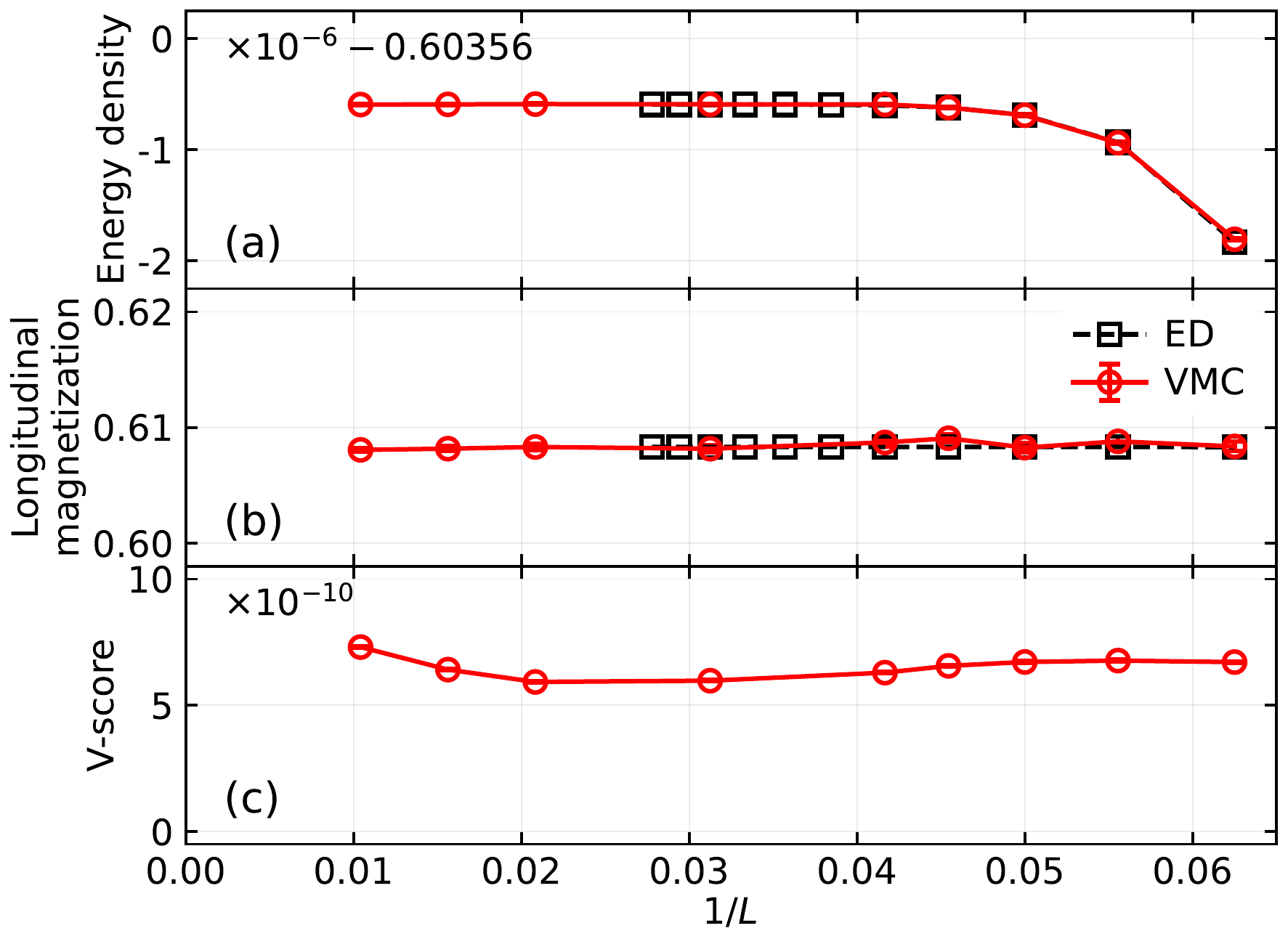}
  \caption{(Color online)
    Numerical results of the 1D pure PXP model.
    We show the energy density (a), the longitudinal magnetization (b), and the V-score (c) as a function of the inverse chain length $1/L=1/N$.
    Red circles are optimized VMC results and black squares are available ED data.
    The error bars are typically smaller than the symbol size.
    The VMC energy and magnetization agree with the ED results within the plotting resolution, while the V-score is of order $10^{-9}$.
  }
  \label{fig:1d}
\end{figure}

Having established the accuracy of the variational state,
we extend the VMC calculations to larger chains beyond the reach of ED.
We summarize the VMC results for the pure PXP chain in Fig.~\ref{fig:1d}.
For all system sizes accessible to ED,
the energy density of the optimized projected BCS state agrees with the exact result within the resolution of the plot.
Its finite-size dependence is weak,
and the energy density rapidly converges toward approximately $-0.60356$.
The longitudinal magnetization is also in agreement with ED
and remains close to $0.608$ over the system sizes that we considered.

Remarkably, the V-score, a more sensitive measure of accuracy, remains of order $10^{-9}$ up to $N=96$.
According to Eq.~\eqref{eq:vscore-energy-error},
this corresponds to an estimated relative energy error of order $10^{-10}$,
although this estimate is subject to the empirical scatter in the V-score benchmark.
For the present calculations,
the estimated relative energy error is comparable to the statistical uncertainty from MC sampling.
Such accuracy is achieved by optimizing all pairing amplitudes for the pure PXP Hamiltonian, which is not frustration free.
The V-score achieved here falls within the range reported for accurate MPS calculations of 1D spin systems~\cite{wu2024},
suggesting a comparable V-score level of accuracy for the present VMC state.

The exceptionally small V-score can be attributed to two key features of our construction.
First, the exactly representable frustration-free state provides an accurate starting point for approximating the ground state of the pure PXP model.
Second, the internal pairing amplitudes can be optimized independently,
providing enough flexibility to correct the remaining configuration-dependent deviations with only $O(N)$ symmetry-reduced parameters.

\subsection{Square lattice}

\begin{figure}[t]
  \centering
  \includegraphics[width=\columnwidth]{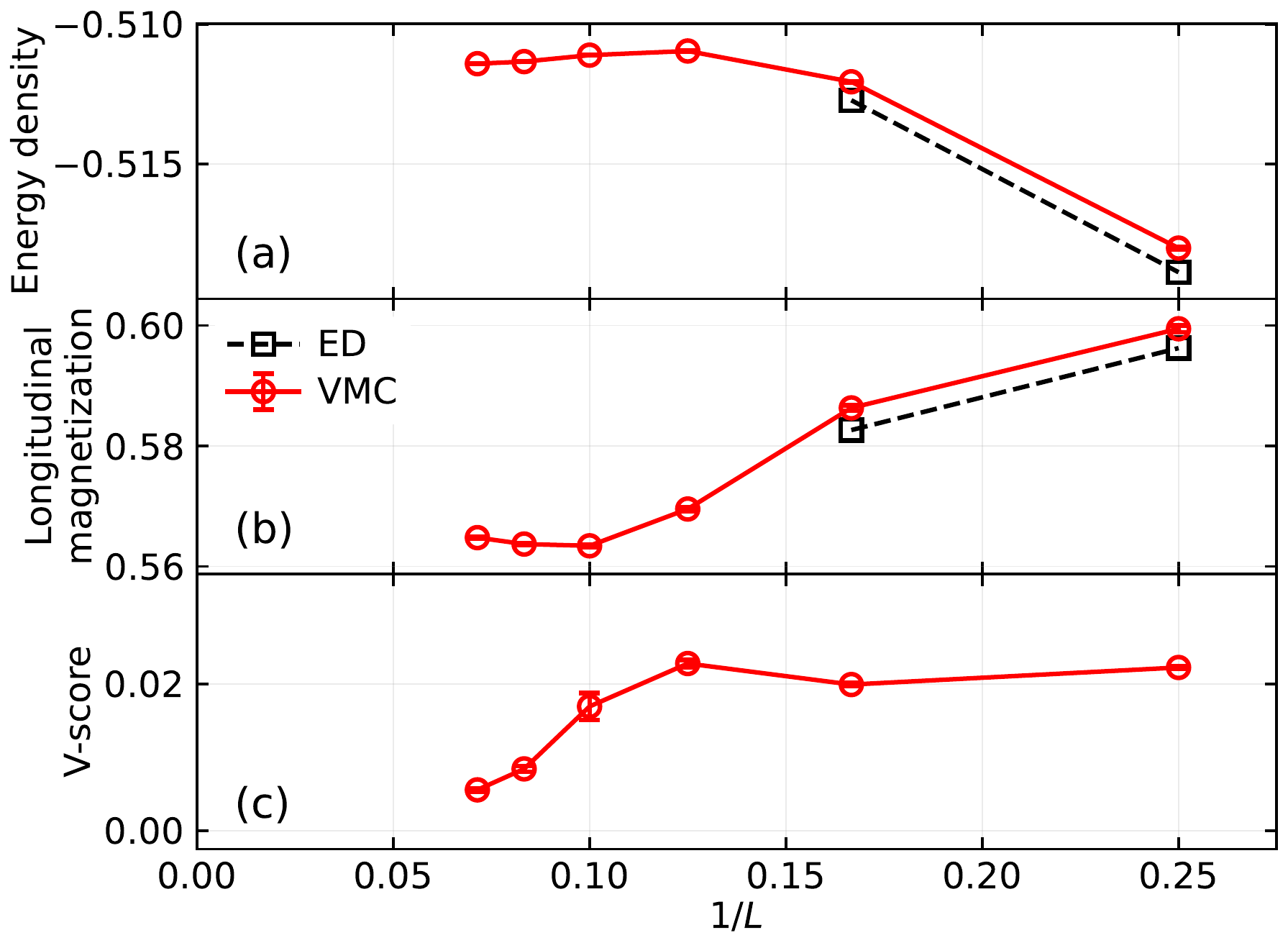}
  \caption{(Color online)
    Numerical results of the square-lattice pure PXP model.
    We show the energy density (a), the longitudinal magnetization (b), and the V-score (c) as a function of the inverse linear size $1/L=1/\sqrt N$.
    Red circles denote VMC results on periodic $N=L^2$ clusters with even $L$, and black squares denote ED results for $L=4,6$.
    The error bars are typically smaller than the symbol size.
    The VMC data extend to larger square clusters beyond the system sizes accessible to ED.
  }
  \label{fig:2d}
\end{figure}

Encouraged by the high accuracy achieved in 1D,
we next apply the same variational approach to the square lattice.
We show the 2D results for the pure PXP model as a function of the inverse linear size $1/\sqrt{N}$ in Fig.~\ref{fig:2d}.
As in 1D, the exactly representable frustration-free state is used only as the initial state,
after which the independent square-lattice pairing amplitudes are optimized variationally.

For comparison with ED, we consider periodic $4\times4$ and $6\times6$ clusters,
for which the relative errors in the variational energies
($\lvert E_{\mathrm{VMC}}-E_{\mathrm{ED}}\rvert/\lvert E_{\mathrm{ED}}\rvert$) are found to be below $0.2\%$.
These clusters are close to the largest sizes accessible to our ED calculations,
given the rapid growth of the constrained Hilbert space.
The VMC calculations can be extended to substantially larger clusters and show a smooth size dependence.
We have confirmed that both the energy density
and the Rydberg-excitation density are consistent with the available continuous-time quantum MC (CTQMC) benchmark.
The energy density appears to approach approximately $-0.511$ with increasing system size,
consistent with the CTQMC estimate of $-0.512$ in the thermodynamic limit~\cite{yue2021}.
The longitudinal magnetization also decreases smoothly and approaches approximately $0.56$.
Using Eq.~\eqref{eq:density-magnetization-conversion},
this corresponds to a Rydberg-excitation density of $n\simeq(1-0.56)/2\simeq0.22$.
Within the present finite-size and variational uncertainties,
this value is consistent with the CTQMC estimate of $n=0.221(5)$ in the thermodynamic limit~\cite{yue2021}.

The V-scores for the calculated system sizes are typically in the range $5\times10^{-3}$ to $2\times10^{-2}$.
According to the empirical relation in Eq.~\eqref{eq:vscore-energy-error},
this range corresponds to estimated relative energy errors of approximately $8\times10^{-4}$ to $3\times10^{-3}$.
Although these errors are larger than those obtained in 1D,
the results show that the projected Pfaffian state still captures the main features of the square-lattice ground state
using only $O(N)$ symmetry-reduced parameters.
At the same time, the larger V-score indicates that the present pairing parameterization does not reproduce the 2D ground state as accurately as it does in 1D.
Nevertheless, it provides a compact and physically motivated description that can be systematically improved by enlarging the variational space.

\section{Conclusions and Outlook}
\label{sec:conclusions}

We developed a fermionic variational description of the PXP ground state
based on a conventional Gutzwiller projector and an explicit microscopic blockade projector.
This construction has three main features.
First, the blockade projector acts directly on physical spin configurations,
independently of the variational Abrikosov-fermion pairing structure.
This differs from the formulations of Refs.~[\citenum{cheng2023,cheng2024}],
which encode the blockade constraint through auxiliary fermions in an enlarged Hilbert space.
Second, the same projected Pfaffian ansatz applies in both one and two dimensions.
Although the pure PXP ground states considered here are expected to belong to
a gapped disordered phase~\cite{li2024} with area-law entanglement~\cite{eisert2010},
tensor-network approaches used for such states generally require dimension-dependent network structures
and become computationally more demanding in higher dimensions.
Our ansatz offers a complementary route in which variational flexibility is controlled
by the number and spatial range of the fermionic pairing amplitudes
rather than by the bond dimension and geometry-dependent connectivity of a tensor network.
Third, the frustration-free deformation provides an exact projected BCS state
that serves as a physically motivated starting point for optimization.
Although this exact construction applies only to the frustration-free Hamiltonian,
optimizing the independent pairing amplitudes yields an accurate description of the non-frustration-free pure PXP ground state,
for which no exact analytical ground-state wave function is known.

In 1D, the optimized state reproduces the ED energies and magnetizations and reaches a V-score of order $10^{-9}$.
Because the V-score can be evaluated without knowing the exact ground state,
it also provides a useful measure of accuracy beyond the sizes accessible to ED~\cite{wu2024}.
The V-scores obtained here are comparable to those typically reported for accurate MPS calculations of 1D spin systems.
On the square lattice, the same ansatz reproduces the available ED and CTQMC results for the energy density and Rydberg-excitation density.
The 2D V-scores are larger than their 1D counterparts,
indicating that the present $O(N)$ parameterization captures the main ground-state properties
while leaving room for further improvement.
Possible extensions include complex pairing amplitudes, additional boundary and gauge sectors,
enlarged unit cells, and backflow or correlator factors~\cite{bajdich2008}.

These results also provide a useful point of comparison with recent neural-network studies of the PXP model~\cite{kufel2025,park2026}.
Neural-network quantum states offer highly flexible representations
and have enabled calculations on large constrained systems,
whereas the present Pfaffian ansatz uses only a compact set of parameters
and is directly motivated by fermionic pairing and the exact frustration-free solution.
Combining these complementary ideas,
for example by augmenting the projected Pfaffian state with a neural-network correlator~\cite{nomura2017,chen2026},
may provide a systematic route to improving the 2D results.
More broadly, the optimized ground states obtained here may provide useful reference states
for exploring excited-state properties and quantum dynamics,
including quantum many-body scars, in constrained systems.
The separation between fermionic single occupancy and the geometrical blockade constraint
should also facilitate extensions to frustrated lattices,
longer-range Rydberg blockade, and constrained quantum magnets in higher dimensions.

\begin{acknowledgments}
This work was financially supported by
MEXT KAKENHI, Grant-in-Aid for Transformative Research Areas
(Grant Nos.\ JP22H05111 and JP22H05114)
and
JSPS KAKENHI
(Grant Nos.\ JP25K07157 and JP26K06959).
The numerical computations were performed on computers at
the Supercomputer Center, Institute for Solid State Physics, University of Tokyo.
\end{acknowledgments}

\balance
\bibliographystyle{jpsjnum}
\bibliography{ref}

\end{document}